\documentclass[letterpaper]{article}

\usepackage{aaai}
\usepackage{color}
\usepackage{times}
\usepackage{helvet}
\usepackage{courier}
\usepackage{graphicx} 
\usepackage{amsmath} 
\usepackage{amssymb}
\usepackage{subfigure}
\usepackage{soul}

\newcommand{\comment}[1]{}
\newcommand{\BEQ}{\begin{equation}}
\newcommand{\EEQ}{\end{equation}}
\newcommand{\BEA}{\begin{eqnarray}}
\newcommand{\EEA}{\end{eqnarray}}
\newcommand{\ssum}{{\sum}}

\renewcommand{\a}{\alpha}
\renewcommand{\b}{\beta}
\newcommand{\yt}{\tilde{y}}

\title{Replicator Dynamics of Co--Evolving Networks\\}
\author{Aram Galstyan \and Ardeshir Kianercy\\
USC Information Sciences Institute\\
Marina del Rey, California 90292, USA\\
\And
Armen Allahverdyan \\ Yerevan Physics Institute \\ Yerevan, Armenia
}
\begin{document} 
\maketitle
\begin{abstract}
\begin{quote}
We propose a simple model of network co--evolution in a game--dynamical system of interacting agents that play repeated games with their neighbors, and  adapt  their behaviors and network links based on the outcome of those games. The adaptation is achieved through a simple reinforcement learning scheme. We show that the collective evolution of such a system can be described by appropriately defined replicator dynamics equations. In particular, we suggest an appropriate factorization of the agents' strategies thats results in a coupled  system of equations characterizing the evolution of both strategies and network structure, and illustrate the framework on two simple examples. 
\end{quote}
\end{abstract}

\section{Introduction}

Many complex systems can be represented as networks where nodes correspond to entities and links encode interdependencies between them.  Generally, statistical models of networks can be classified into two different approaches. In the first approach, networks are modeled via active nodes with a given distribution of links, where each node of the network represents a dynamical system. In this settings, one usually studies problems related to epidemic spreading, opinion formation, signaling and synchronization and so on. In the second approach,  which is grounded mainly in a graph-theoretical approach, nodes are treated as passive elements, Instead, the main focus is on dynamics of link formation and network growth. Specifically, one is interested in  algorithmic methods to build graphs formed by passive elements
(nodes) and links, which evolve according to pre-specified, often local rules. This approach produced important results on topological features
of social, technological and biological networks.

More recently, however, it has been realized that modeling individual and network dynamics separately is too limited to capture realistic behavior of networks. Indeed, most real--world networks are inherently complex dynamical systems, where both  attributes of individuals (nodes) and topology of the network (links) can have inter--coupled dynamics.  For instance, it is known that in social networks, nodes tend to divide into groups, or communities, of like-minded individuals. One can ask   whether individuals become likeminded because they
are connected via the network, or whether they form network connections
because they are like-minded.  Clearly, the distinction between the two scenarios is not clear-cut. Rather, the real world
self-organizes by a combination of the two, the network changing in
response to opinion and opinion changing in response to the network.   Recent research has focused on the  interplay between attribute and link  dynamics (e.g., see~\cite{Gross2008,goyal2005learning,perc2009coevolutionary,Fortunato2009} for a recent survey of the literature).

To describe coupled  dynamics of individual attributes and network topology, here we suggest a simple model of co--evolving network that is based on the notion of interacting adaptive agents. Specifically,  we consider network--augmented multi--agent systems where agents play repeated game with their neighbors, and  adapt both their behaviors and the network ties depending on the outcome of their  interactions. To adapt,  agents use a simple learning mechanism to reinforce (punish) behaviors and  network links that produce favorable (unfavorable) outcomes. We show that the collective evolution of such a system can be described by appropriately defined replicator dynamics equations. Originally suggested in the context of evolutionary game theory (e.g., see~\cite{Hofbauer1998Book,hofbauer2003evolutionary}), replicator equations have been used  to model collective learning and adaptation in a systems of interacting self--interested agents~\cite{Sato2003}.

\section{Background and Related Work}
One of the oldest and best studied models of a network is the  Erd\"{o}s--R\'{e}nyi random graph  defined as $G(N;p)$ where $N$ is the number of vertices, and $p$ is the probability of a link between any two vertices. One of the important topological features of graphs is the degree distribution $p_k$, which is the probability that a randomly chosen node has exactly $k$ neighbors. In the large $N$ limit, the degree distribution of the Erd\"{o}s--R\'{e}nyi graph  is Poissonian, $p_k=e^{-z}z^k/k!$, where $z=pN$ is the average {\em degree}, or connectivity.  While this 
model adequately describes the percolation transition of the real
networks, it fails to account for many properties of real--word networks  such as the
Internet, social networks or biological networks. In particular, it has been established that many real--world network exhibit what is called a scale--free phenomenon, where the degree distribution follows a power law $p_k \sim k^{-\gamma}$ over a very wide (orders of magnitude) range of $k$. 

To account for the statistical deviations of the observed properties of networks from those prescribed by the  Erd\"{o}s--R\'{e}nyi random graph model,   Barabasi and Albert~\cite{Barabasi1999} proposed a simple model of an evolving network, based on an idea of preferential attachment. In this model, a network grows by addition of new nodes at each time step. Each new node introduced in the system chooses to connect {\em preferentially} to sites that are already well connected.  Thus, nodes that have higher connectivity will add new links with higher rates.  It was shown that the network produced by this simple process has an asymptotic scale--free degree distribution of form $p_k\sim k^{-3}$. Recent variations of the initial preferential attachment model include
space-inhomogeneous~\cite{Bianconi2001} and time-inhomogeneous 
generalizations of the preferential attachment mechanism \cite{Dorogovtsev2001}, ageing and redistribution of the 
existing links~\cite{Dorogovtsev2000}, preferential attachment
with memory~\cite{Cattuto2006}, evolutionary generalizations of the preferential attachment
\cite{Poncela2008}, {\it etc}.

\cite{Holme2006} suggested a  model co--evolving networks that combines linking with internal node dynamics. In their model, each node is assumed to hold one of $M$ possible opinions. Initially, the links are randomly distributed among the nodes. Then, at each time step,  a randomly chosen node will re--link, with probability $\phi$, one of his links to a node that holds the same opinion. And with probability $1-\phi$, he will change his opinion to agree with the opinion of one of his (randomly chosen) neighbor. Despite the simplicity of those rules, the model was shown to have a very rich dynamical behavior. In particular, while varying the parameter $\phi$, the model undergoes a  phase transition from a
phase in which opinions are diverse to one in which most individuals hold the same opinion. \cite{Skyrms2000} suggested a model of adaptive networks where agents are allowed to interact with each other through games, and reinforce links positively  if the outcome of the interaction is favorable to them. They showed that even for the simple games, the resulting structural dynamics can be very complex. A review of numerous other models can be found in a recent survey~\cite{Fortunato2009}.

In addition to  abstract statistical models, recent work has addressed the network formation process from the perspective of game--theoretical interactions between self--interested agents~\cite{Bala2000,Fabrikant2003,Anshelevich2003}. In these games each agent tries to maximize her utility consisted of two conflicting preferences -- e.g., minimizing the cost incurred by established edges, and  minimizing the distance from all the other nodes in the networks. In the simplest scenarios of those games, the topology that corresponds to the Nash equilibrium can be obtained by solving a one--shot optimization problem. In many situations, however, when the actual cost function is more complex, this might not be possible. Furthermore, in realistic situations, agents might have only local information about the network's topology (or utilities of the other agents in the network), so maximizing a global utility is not an option.  In this case, the agents can arrive at Nash equilibrium by dynamically adjusting their strategies.

\section{Dynamics for Co-Evolving Networks }

Let us  consider a set of agents that play repeated games with each other. Each round of  the game proceeds as follows: First, an agent has to choose what other agent he wants to play with. Then, assuming that the other agent has {\em agreed} to play, the agent has to choose an appropriate action from the pool of  available actions. Thus,  to define an overall game  strategy, we have to specify how  an agent chooses a partner for the game and a  particular action.  

For the sake of simplicity, let us start with three agents, which is the minimum number required for a non--trivial dynamics. Let us differentiate those agents by indices $x$, $y$, and $z$.  Here we will focus on the case when the number of actions available to agents is finite. The time--dependent mixed strategies of agents can be characterized by a probability distribution over the choice of the neighbors and the actions.  For instance, $p_{xy}^i(t)$ is the probability that the agent $x$ will choose to play with agent $y$ and perform action $i$ at time $t$.

Furthermore, we assume that the agent adapt to their environment through a simple reinforcement mechanism. Among different reinforcement schemes, here we focus on (stateless) $Q$-learning~\cite{Watkins1992}. Within this scheme, the  agents' strategies are parameterized through so called $Q$--functions that characterize relative utility of a particular strategy. After each round of game, the $Q$ functions are updated according to the following rule:  
\BEQ
\label{ararat}
Q_{xy}^i(t+1) = Q_{xy}^i(t) + \a [R_{x,y}^i -  Q_{xy}^i(t)  ] 
\EEQ
where $R_{x,y}^i$ is the expected reward of agent $x$ for playing action $i$ with  agent $y$, and $\a$ is a parameter that determines the learning rate (which can be set to $\a=1$ without a loss of generality). 

Next, we have to specify how agents choose a particular neighbor and an action based on their $Q$-function.  Here we use the Boltzmann exploration mechanism  where the probability of a particular choice is given as~\cite{Sutton_book}
\BEQ
\label{dynamo}
p_{xy}^i = \frac{e^{\b Q_{xy}^i}}{\sum_{\yt,j}e^{\b Q_{x\yt}^j}}
\EEQ
Here  the inverse {\em temperature} $\b=1/T>0$ controls exploration/exploitation tradeoff: for $T\rightarrow  0$ the agent always choose the action corresponding to the maximum $Q$--value, while for $T\rightarrow \infty$ the agents' choices are completely random. 

We now assume that the agents interact with each other many times between two consecutive updates of their strategies. In this case, the reward of the $i$--th agent in Equation~\ref{ararat} should be understood in terms of the {\em average reward}, where the average is taken  over the strategies of other agents, $R_{x,y}^i = \sum_{j} A^{i,j}_{xy} p_{y x}^j$, where $A^{i,j}_{xy}$ is the reward (payoff) of agent $x$ playing strategy $i$ against the agent $y$ who plays strategy $j$. Note that generally speaking, the payoff  might be asymmetric.

We are interested in the continuous approximation to the learning dynamics.  Thus, we replace $t+1 \rightarrow t+\delta t$, $\alpha \rightarrow \alpha \delta t$, and take the limit $\delta t \rightarrow 0$ in (\ref{ararat}) to obtain 
\BEQ
\label{ararat2}
\dot{Q}_{xy}^i = \a [R_{x,y}^i -  Q_{xy}^i(t)  ] 
\EEQ
Differentiating \ref{dynamo}, using Eqs.~\ref{dynamo},~\ref{ararat2}, and scaling the time  $t\rightarrow \a\b t$  we obtain the following replicator equation~\cite{Sato2003}:
\BEQ
\label{real}
\frac{ \dot{p}_{xy}^i}{p_{xy}^i} =    \sum_{j} A^{i,j}_{xy} p_{y x}^j   -  \sum_{i,j,\yt} A^{i,j}_{x,\yt} p_{x \yt}^i  p_{\yt x}^j  + T  \sum_{\yt,j} p_{x\yt}^j \ln \frac{p_{x\yt}^j}{p_{xy}^i}
\EEQ
Equations~\ref{real} describes the collective adaptation of the Q--learning agents through repeated game--dynamical interactions. The first two terms indicate  that a probability of a playing a particular pure strategy increases with a rate proportional to the overall goodness of that strategy, which mimics fitness-based selection mechanism in population biology~\cite{Hofbauer1998Book}. The second term, which has an entropic meaning,   does not have a direct analogue in population biology~\cite{Sato2003}. This term is due to the Boltzmann selection mechanism, and thus, describes the agents' tendency to randomize over their strategies. Note that for $T=0$  this term disappears and the equations reduce to the conventional replicator system~\cite{Hofbauer1998Book}. 

So far, our discussion has been very general. We now make the assumption that the agents strategies can be factorized   as follows:
\BEQ
\label{barca}
p_{xy}^i = c_{xy} p_{x}^i \ , \ \ssum_{y} c_{xy} = 1, \  \ \ssum_{i} p_{x}^i = 1 .
\EEQ
Here $c_{xy}$ is the probability that the agent $x$ will initiate a  game with the agent $y$, whereas $p_x^i$ is the probability that he will choose action $i$.  Thus, the assumption behind this factorization is that  the probability that the agent will perform action $i$ does not depend on whom the game is played against. 

To proceed further, we substitute~\ref{barca} in~\ref{real},  take a summation of both sides in the above equation once  over $y$ and then over $i$, and make use of the normalization conditions in Eq.~\ref{barca} to obtain the following system: 

\BEA
\frac{\dot{p}_{x}^i}{p_x^i} &=&  \sum_{\yt,j} A^{ij}_{x\yt} c_{x\yt}c_{\yt x}p_{\yt}^j  -  \sum_{i,j,\yt} A^{ij}_{x\yt} c_{x\yt}c_{\yt x}p_x^i p_{\yt}^j   \nonumber \\
&+&T\sum_j p_x^j \ln( p_x^j/p_x^i )
\label{sevilia1}\\
  \frac{\dot{c}_{xy}}{c_{xy}}  &=&  c_{yx}\sum_{i,j} A^{ij}_{xy} p_x^i p_y^j  - \sum_{i,j,\yt} A^{ij}_{x\yt} c_{x\yt}c_{\yt x}p_x^i p_{\yt}^j   \nonumber \\ 
&+&  T\sum_{\yt} c_{x \yt} \ln( c_{x \yt}/c_{xy} )\label{sevilia2}
\EEA

 Equations~\ref{sevilia1} and ~\ref{sevilia2} are the replicator equations that describe the collective and mutual evolution of the agent strategies and the network structure, by taking into account explicit coupling between the strategies and link wights.  Our preliminary  analysis suggest that this co--evolutionary system  can demonstrate a very rich behavior even for  simple games. Below we illustrate the framework on two simple examples.

\subsection{Examples}
Our preliminary results indicate that the co--evolutionary system Equations~\ref{sevilia1} and~\ref{sevilia2}  can have a very rich behavior even for  simple games. The full analysis of those equations will be reported elsewhere. Here we consider  two simple examples, and focus on the link dynamics (Eqs.~\ref{sevilia2}), assuming that  the agents play Nash equilibrium strategies of the corresponding two--agent game. 

Our first example is a  simple coordination game with the following (two--agent) payoff matrix:
 \[
A=\left(
\begin{array}{cc}
 1 &0   \\
  0&0\\
\end{array}
\right)
\]
Thus, agents get a unit reward if they jointly  take the first action, and get no reward otherwise.

\begin{figure}[!htb]
\centering
\subfigure[]{
    \includegraphics[width = 0.33\textwidth ]{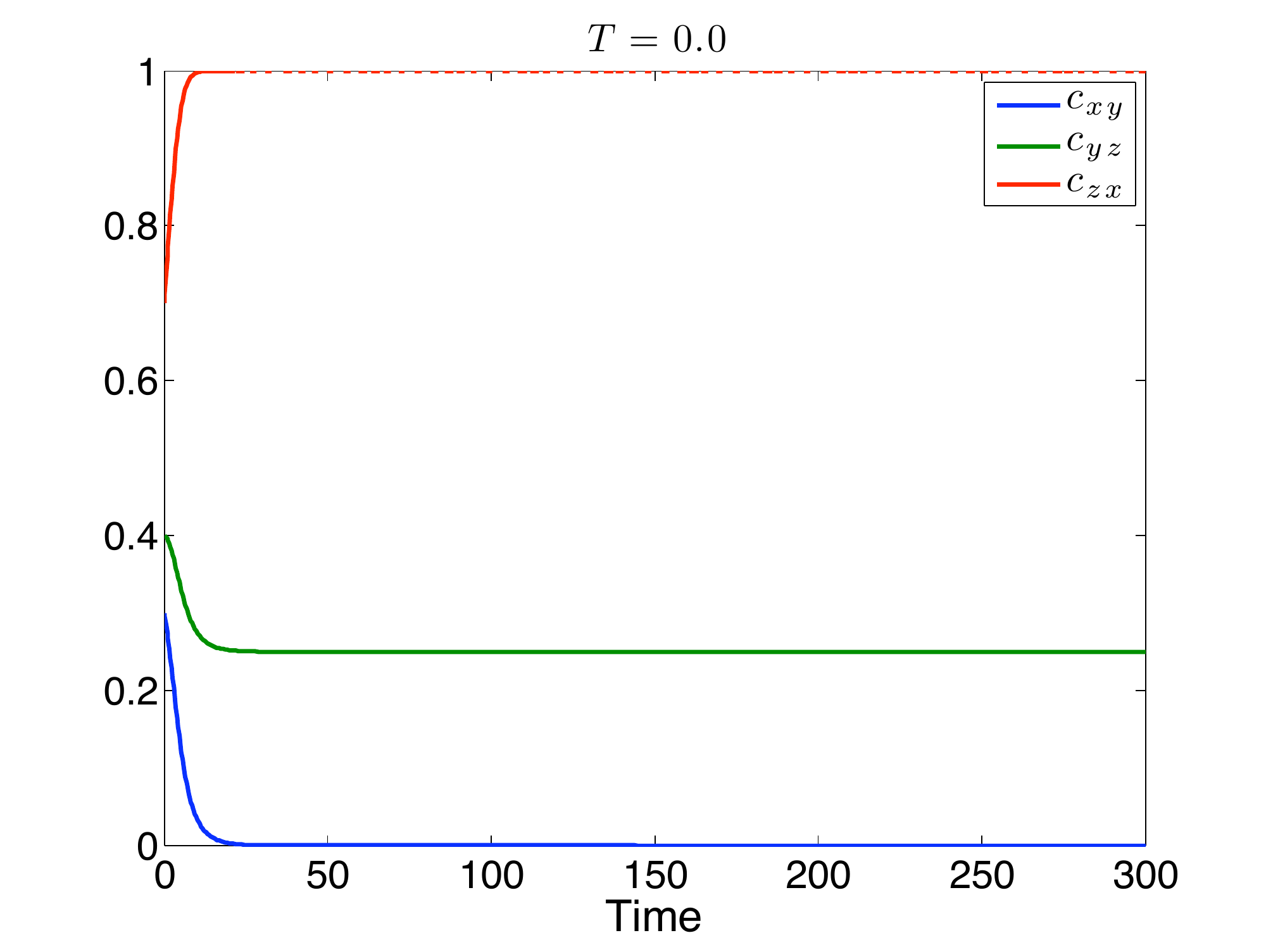} \label{fig1a}
    } 
    \subfigure[]{
    \includegraphics[width = 0.33\textwidth]{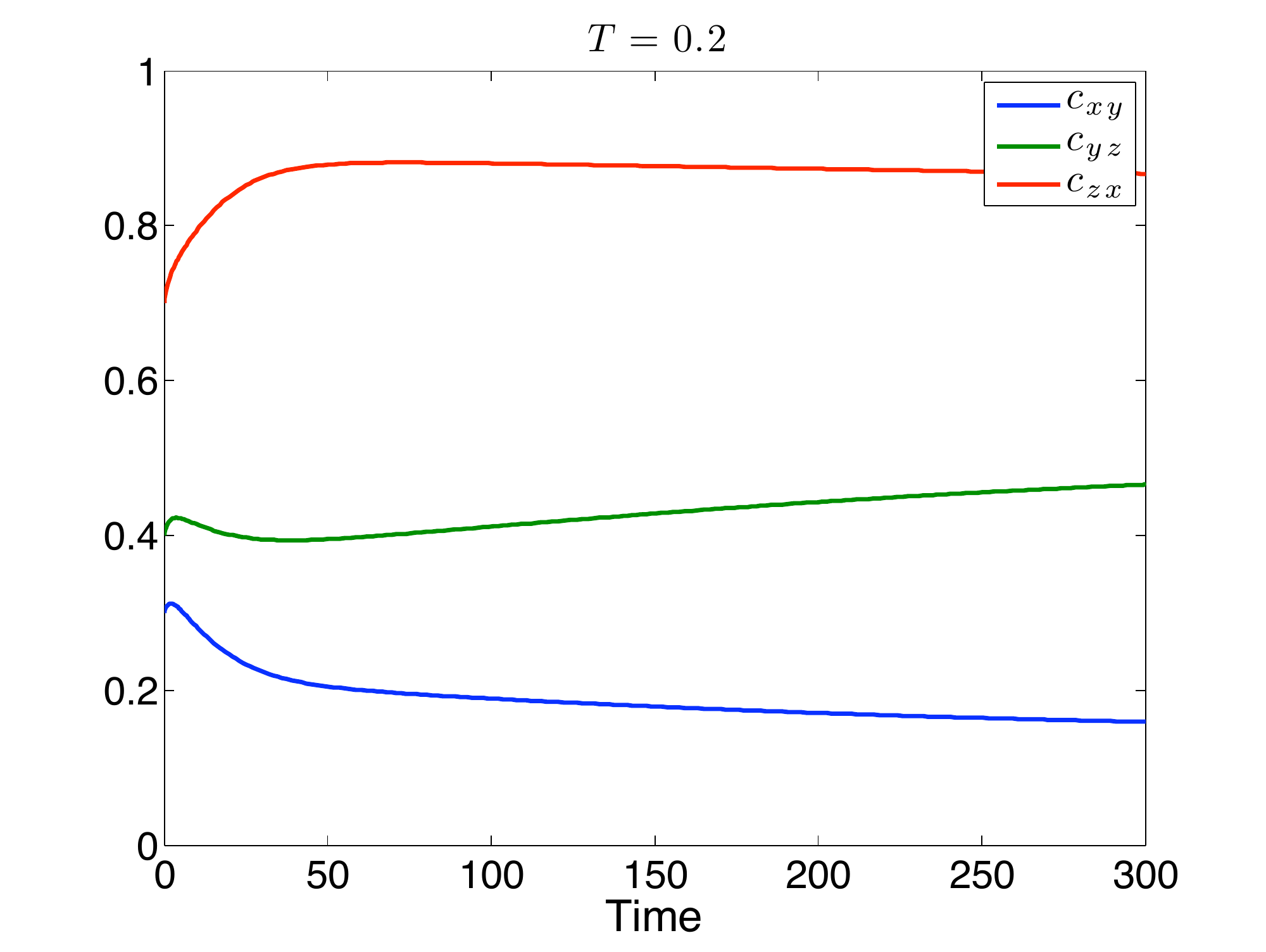} \label{fig1b}
    }
    \subfigure[]{
    \includegraphics[width = 0.33\textwidth]{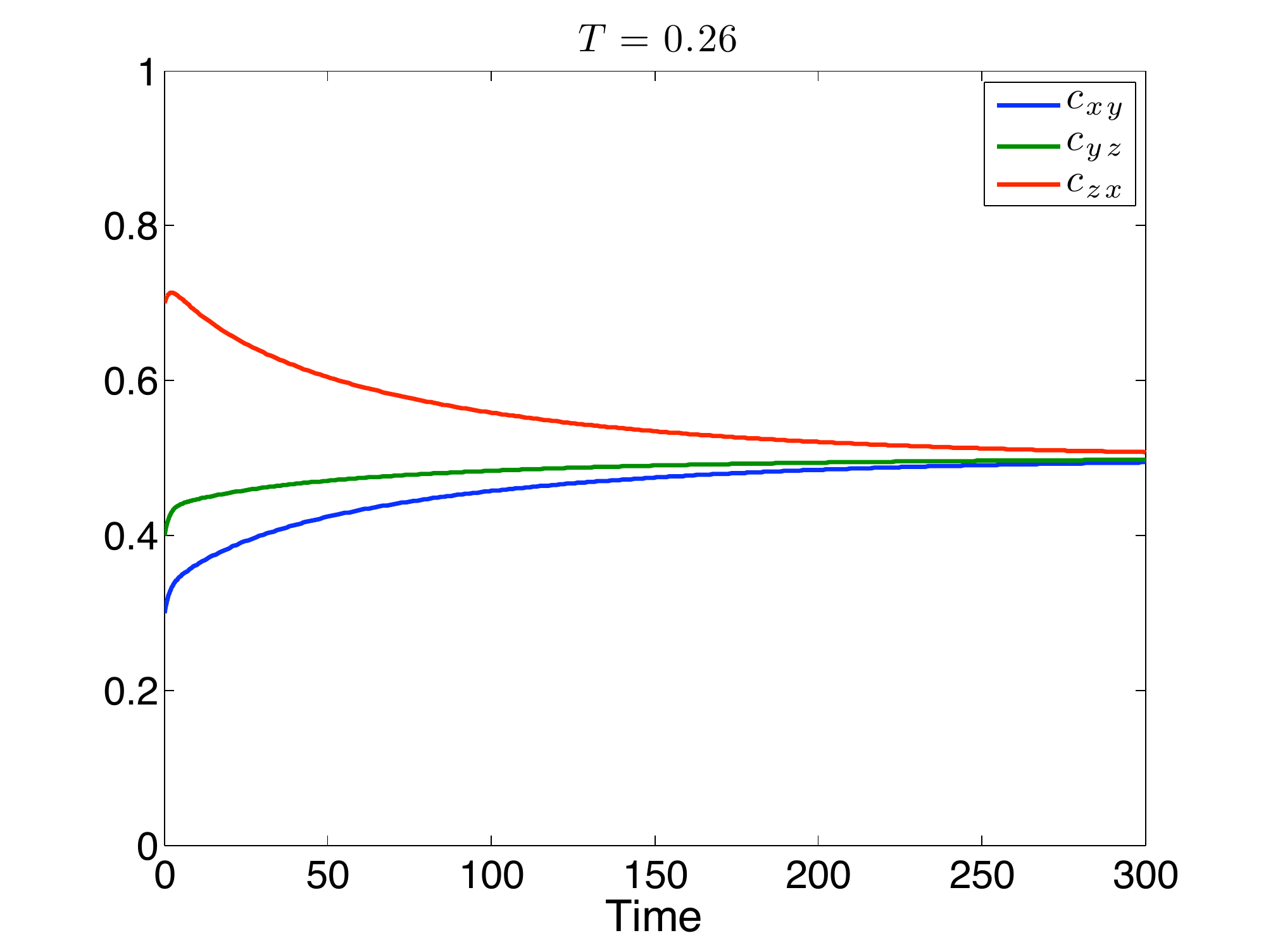} \label{fig1c}
    }
     \caption{Dynamics of links for various temperatures. }
     \label{fig:1}
\end{figure}

We assume that the agents always play the same action (e.g., $p_x^1=p_y^1=p_z^1\equiv1$) yielding a reward $A^{11}=1$,  so we can focus on the link dynamics.   Then the   equations characterizing the link dynamics are as follows: 
\BEA
\frac{ \dot{c}_{xy}}{c_{xy}(1-c_{xy})}  &=&  1 - c_{yz} - c_{zx} + T\ln\frac{1-c_{xy}}{c_{xy}} 
\label{betis4}\\
\frac{ \dot{c}_{yz}}{c_{yz}(1-c_{yz})} &=&  1 - c_{zx}  - c_{xy} + T\ln\frac{1-c_{yz}}{c_{yz}}
\label{betis5}\\
 \frac{\dot{c}_{zx}}{c_{zx}(1-c_{zx})} &=& 1 -c_{xy} - c_{yz}  +T\ln\frac{1-c_{zx}}{c_{zx}}
\label{betis6}
\EEA
We note that the system allows different rest--points some of which correspond to pure Nash equilibria (NE). For instance, one such configuration is $c_{xy}=1-c_{yz}=1$, while $c_{zx}$ can have arbitrary value. In this NE configuration agents $x$ and $y$ always play against each other while agent $z$ is isolated.  In addition, there is an interior rest point at $c_{xy}=c_{yz}=c_{zx}=1/2$, which is again a NE configuration. A simple analysis yields that this symmetric  rest point is unstable if the temperature is below a certain critical value. This can be shown by linearizing the system around the symmetric rest point and obtaining the following Jacobian: 
 \begin{eqnarray}
J_T= \frac{1}{4}\left( \begin{array}{ccc} -4T & -1 & -1 \\ -1 & -4T & -1 \\ -1 & -1 & -4T  \end{array} \right), 
  \end{eqnarray}
 It is straightforward to show that for $4T>1$ all three eigenvalues of this Jacobian become negative, thus making the interior rest point stable. The dynamic of links for various temperature is shown in Figure~\ref{fig:1}.  


As a second example, we consider the Rock-Paper-Scissor (RPS) normal game which has the following payoff matrix:
 \[
A=\left(
\begin{array}{ccc}
 \epsilon&-1&1   \\
  1&\epsilon&-1\\
  -1&1& \epsilon   
\end{array}
\right)
\]
where $-1 < \epsilon < 1$. This game has a mixed Nash equilibrium where all the strategies are played with equal probabilities $1/3$. Note that RPS game can have a very rich and interesting dynamics even for two players.  For instance, it has been noted that for a two-person RPS game  at $T=0$ the dynamical system might show a chaotic behavior at certain range of $ \epsilon$ and never reach an equilibrium~\cite{sato2002chaos}. Again, here we focus on the link dynamics by assuming that the agents play the NE--prescribed strategies $p_x^i=p_y^i=p_z^i=\frac{1}{3}, i=1,2,3$. The equations describing the evolution of the links are as follows:
\BEA
\frac{ \dot{c}_{xy}}{c_{xy}(1-c_{xy})}  &=& \frac{\epsilon_{x}}{3} (1 - c_{yz} - c_{zx}) + T\ln\frac{1-c_{xy}}{c_{xy}} \nonumber
\label{bbetis15}\\
\frac{ \dot{c}_{yz}}{c_{yz}(1-c_{yz})} &=&  \frac{\epsilon_{y}}{3} ( 1 - c_{zx}  - c_{xy} )+ T\ln\frac{1-c_{yz}}{c_{yz}}\nonumber
\label{bbetis16}\\
 \frac{\dot{c}_{zx}}{c_{zx}(1-c_{zx})} &=&\frac{\epsilon_{z}}{3} ( 1 -c_{xy} - c_{yz} ) +T\ln\frac{1-c_{zx}}{c_{zx}}\nonumber
 \label{bbetis17}
\EEA
This system has a number of different rest points. For instance, for $-1<\epsilon<0$ and $T=0$, the stable rest point corresponds to a directed triangle with no reciprocated links, $c_{xy}=c_{yz}=c_{zx}=1$ or  $c_{xy}=c_{yz}=c_{zx}=0$ . This is expected, since for $-1<\epsilon < 0 $, the average reward is negative, so the agents are better off not playing with each other at all. There is also an interior rest point at $c_{xy}=c_{yz}=c_{zx}=\frac{1}{2}$. As in the previous example, it can be shown that there is a critical value of $T$ below which this rest point is unstable. Indeed, the Jacobian around the symmetric rest point is as follows: 
\begin{equation}J=\frac{-1}{12}\left(\begin{array}{ccc} 12\, T & \mathrm{\epsilon} & \mathrm{\epsilon}\\ \mathrm{\epsilon} & 12\, T & \mathrm{\epsilon}\\ \mathrm{\epsilon} & \mathrm{\epsilon} & 12\, T \end{array}\right)
\end{equation}
A simple calculation shows that the interior rest point becomes stable whenever $T>\epsilon/12$ for $1>\epsilon > 0$, and $T>|\epsilon|/6$ for $-1<\epsilon < 0$.

\section{Discussion}

In conclusion, we have presented a replicator--dynamics based framework for studying mutual evolution network topology and agent behavior in a network--augmented system of interacting  adaptive agents. By assuming that the agents strategies allow appropriate factorization, we derived a system of a coupled replicator equations  that describe the mutual evolution of agent behavior and network link structure.  The examples analyzed here were for simplified scenarios.   As a future work, we plan to perform a more through analysis of the dynamics for the fully coupled system. Furthermore, we intend to go beyond the three--agent systems considered here and examine larger systems. Finally, we note that the main premise behind our model is that the strategies can be factorized according to Equations~\ref{barca}. While this assumption is justified for certain type of  games, this might not be the case for  some other games, where factorization can destroy some NE points that exist in   the non--factorized strategy spaces. We intend to examine this question more thoroughly in our future work.   

 \section{Acknowledgments}
This research was supported by the National Science Foundation under grant No. 0916534.

\bibliographystyle{aaai}


\end{document}